\documentclass[5p,twocolumn]{elsarticle}
\usepackage{graphicx,latexsym}
\usepackage{dcolumn}
\usepackage{amssymb,amsmath,bm}
\usepackage{subfigure}
\usepackage{braket}
\usepackage{siunitx}
\usepackage{array}
\usepackage{float}
\usepackage[nopar]{lipsum}
\usepackage{caption}
\usepackage{multirow}
\usepackage{bigstrut}
\usepackage[super]{nth}

\newcommand{\angstrom}{\text{\normalfont\AA}}
\usepackage{xurl}
\usepackage{hyperref}
\hypersetup{
    pdfnewwindow=true,       
    colorlinks=true,         
    linkcolor=blue,          
    citecolor=blue,          
    filecolor=magenta,       
    urlcolor=black           
}

\usepackage[normalem]{ulem}

\def\sec#1{Sec.\ \ref{#1}}

\def\fig#1{Fig.\ \ref{#1}}
\def\tab#1{Tab.\ \ref{#1}}

\journal{}

\begin{document}

\begin{frontmatter}


\title{Modulation of electronic and thermal proprieties of TaMoS$_2$ by controlling\break 
	the repulsive interaction between Ta dopant atoms}

\author[a1,a2]{Nzar Rauf Abdullah}
\ead{nzar.r.abdullah@gmail.com}
\address[a1]{Division of Computational Nanoscience, Physics Department, College of Science, 
	University of Sulaimani, Sulaimani 46001, Kurdistan Region, Iraq}
\address[a2]{Computer Engineering Department, College of Engineering, Komar University of Science and Technology, Sulaimani 46001, Kurdistan Region, Iraq}

\author[a3]{Botan Jawdat Abdullah}
\address[a3]{Physics Department, College of Science- Physics- Salahaddin University-Erbil}

\author[a1]{Hunar Omar Rashid}

\author[a4]{Chi-Shung Tang}
\address[a4]{Department of Mechanical Engineering,
	National United University, 1, Lienda, Miaoli 36003, Taiwan}   

\author[a5]{Vidar Gudmundsson}
\address[a5]{Science Institute, University of Iceland, Dunhaga 3, IS-107 Reykjavik, Iceland}


\begin{abstract}

We theoretically study the electronic and the thermal characteristics of Tantalum, Ta, doped Molybdenum disulfide, MoS$_2$, using density functional theory. It has been shown that the MoS$_2$ monolayer is not a good material for thermoelectric devices due to its relatively large band gap. 
We find that a Ta doped MoS$_2$ forming a TaMoS$_2$ monolayer can be useful for thermoelectric devices. The particular attention of this work is paid to the interaction effect between the Ta atoms in the MoS$_2$ structure. We find that the interaction type is repulsive. It introduces an asymmetry in the density of states, DOS, 
reducing the band gap. 
In the presence of a strong repulsive interaction of Ta-Ta atoms, new states in the DOS around the Fermi energy are found leading to a reduction of the band gap. 
Consequently, a high Seebeck coefficient and figure of merit are seen over a wide range of energy around the Fermi energy. In contrast, a small reduction of the band gap and a vanishing degeneracy of the valence and the conduction bands are observed for the case of a weak Ta-Ta repulsive interaction leading to less promising thermoelectric properties.   
\end{abstract}

\begin{keyword}
Thermoelectric \sep MoS$_2$ \sep DFT \sep Electronic structure,  \sep Doping 
\end{keyword}

\end{frontmatter}

Two-dimensional (2D) transition metal dichalcogenides (TMDs) materials are attracting a significant deal of interest due to their unique properties and great potential for applications in nanotechnology \cite{Wang2016, Kim_2020, Asokan2018}.  Among all 2D TMDs, Molybdenum disulfide (MoS$_2$) has turned out to be one of the most interesting materials for applications due to a series of special and tunable physical and chemical properties \cite{C4RA11852A, mi12030240, ABDULLAH2020126807} in various fields with potential as field-effect transistors \cite{doi:10.1021/nn404961e}, solar energy devices \cite{doi:10.1021/nn303772b}, energy conversion \cite{doi:10.1021/nn5013429} and energy storage devices \cite{C1NR11552A}, photoluminescence \cite{doi:10.1021/nl903868w}, catalysis for hydrogen evolution reaction \cite{doi:10.1021/ja201269b}, piezoelectric devices \cite{Wu2014}, photo-electro catalysis \cite{C3CY00207A, doi:10.1021/acscatal.7b01517}  electrical-thermal conductivity \cite{Lee2016, ABDULLAH2021114644} and sensor technologies \cite{10.3389/fchem.2019.00708, s18113638}. 

More significantly, a monolayer MoS$_2$ has a large direct band gap of $1.8$ eV, whereas a bulk crystal has an indirect gap semiconductor with a band gap of $1.29$ eV \cite{PhysRevLett.105.136805, PhysRevB.57.6666}. Therefore, 2D MoS$_2$ has a superior electric performance calling for more attention among the 2D TMDs materials.
Doping nanomaterials provides an additional flexibility to change their characteristics. Lately, the dependence of the physical properties on the system size have been studied experimentally and theoretically for both pure and doped form of 2D materials as it provides a flexible way to alter the properties of materials for different applications. Transition elements have been used for doping 2D-TMDCs.

There are several recent theoretical studies predicting specific properties of MoS$_2$ monolayers with
different dopants. A MoS$_2$ monolayer with doped Re and Nb can lead to both n-type and p-type conductivity according to DFT-LSDA calculations \cite{PhysRevB.88.075420}. A DFT-LDA modeling shows that the dielectric 
environment of a MoS$_2$ monolayer affects the transition atom states leading to shallow energy levels \cite{PhysRevB.92.115431}. The Re and Au dopants have been shown to be dispersed or aggregated in the MoS$_2$ layers providing catalytically active sites within a spin polarized DFT, Ti and V dopants enhance the electrical conductivity and photosensitivity of MoS$_2$ according to a DFT-GGA calculation \cite{WILLIAMSON2017157, ABDULLAH2020126350}. DFT calculations of the change in the Gibbs free energy in relation to hydrogen adsorption indicate that the basal plane of the distorted tetragonal structured alloy catalyst becomes active for hydrogen evolution reaction when alloyed with Ra \cite{https://doi.org/10.1002/adma.201803477}. DFT calculations at the GGA level predict that a MoS$_2$ monolayer is nonmagnetic and is energy favorable for ferromagnetic coupling with Co dopants inducing spin polarized states around the Fermi energy \cite{Wang2016}. Ni doped MoS$_2$ monolayer has been proposed as a novel gas adsorbent to remove the typical decomposition components of H$_2$S and SO$_2$ molecules interacting with the Ni-MoS$_2$ surface based on DFT-GGA calculation \cite{nano8090646}. The effects of V dopants on the antiferromagnetic and the ferromagnetic states of a monolayer MoS$_2$ depend on the separation between the V dopants was found in a DFT-GGA calculation\cite{MIAO2018226}. Furthermore, Mn, Fe, Co, V, Nb and Ta have been used as dopants in MoS$_2$ monolayers to investigate the electronic properties within a DFT at the GGA level and the results indicated that the impurity elements have very low formation energies for their substitution of a Mo atom achieving the most negative value for Ta \cite{Lu2014}. Choi has shown how strain changes electronic properties of p doped MoS$_2$ with Nb and Ta through first-principles hybrid functional calculations \cite{PhysRevApplied.9.024009}.

In this work, a monolayer MoS$_2$ doped with Ta is investigated and the obtained results of the 
DFT-GGA calculations are analyzed to further predict the electronic and thermal properties \cite{ABDULLAH2020114556}. The effects of the repulsive interaction between the dopant Ta atoms 
in the MoS$_2$ monolayer are considered and how they can be used to tune the electronic and thermal properties.
This tuning possibility has not reported previously. Especially, we find a high figure of merit, ZT, due to a change in the electrical conductivity, thermal conductivity and Seebeck coefficient \cite{RASHID2019102625}. This study reveals that Ta doping of MoS$_2$ can have a novel impact on the electronic and thermal properties of MoS$_2$ monolayers. 

In \sec{Sec:Model} the computational tools are shown. In \sec{Sec:Results} the main achieved results are analyzed. In \sec{Sec:Conclusion} the conclusion of results is presented.

\section{Computational Tools}\label{Sec:Model}

To model and visualize the structures under investigation, we use XCrySDen and VESTA \cite{KOKALJ1999176, momma2011vesta}. The density functional theory (DFT) implemented in Quantum espresso (QE) package is used to study the physical properties of the systems \cite{Giannozzi_2009, giannozzi2017advanced}. 
A projector augmented wave (PAW)
potential is utilized to treat the electron–ion interactions and the valence electrons are evaluated explicitly with a plane-wave basis set with a cutoff energy of 1088.45 eV \cite{ABDULLAH2020100740}. The Perdew-Burke-Ernzerhof (PBE) functional is used to calculate the exchange-correlation energy.
The structures are considered fully relaxed on a $14\times14\times1$ $k$-mesh grid when the changes in energy and forces are less than $10^{-8}$ eV and $10^{-5}$ eV/$\angstrom$, respectively \cite{ABDULLAH2020100740}.

A convergence criteria for the energy of the Self-Consistent Field (SCF) is tested on the $14\times14\times1$ 
k-mesh grid, but for the density of states (DOS) calculation a $100\times100\times1$ k-mesh is utilized.
All the structures are simulated with spacers of $20 \, \angstrom$ of vacuum on either side of the 
2d-structure in order to avoid interaction between charge images in the sheets.

Finally, to calculate thermometric characteristics, the Boltzmann transport properties software package (BoltzTraP) is used \cite{Madsen2006}. The BoltzTraP code uses a mesh of band energies and has an interface to the QE package \cite{ABDULLAH2021106981, ABDULLAH2021110095}.  

\section{Results}\label{Sec:Results}

We consider a $2\times2$ supercell for the MoS$_2$ and the Ta doped MoS$_2$ monolayers. 
The Ta atoms in MoS$_2$ do energetically interact with each other, and the interaction effects influence the physical properties of the material.
The analysis of the effects of the Ta-Ta interactions on the TaMoS$_2$ monolayers must be based on a comparison to
a pristine MoS$_2$ monolayer. Because of that we briefly revise the electronic and the thermal properties of a MoS$_2$ monolayer.

\begin{figure}[htb]
	\centering
	\includegraphics[width=0.5\textwidth]{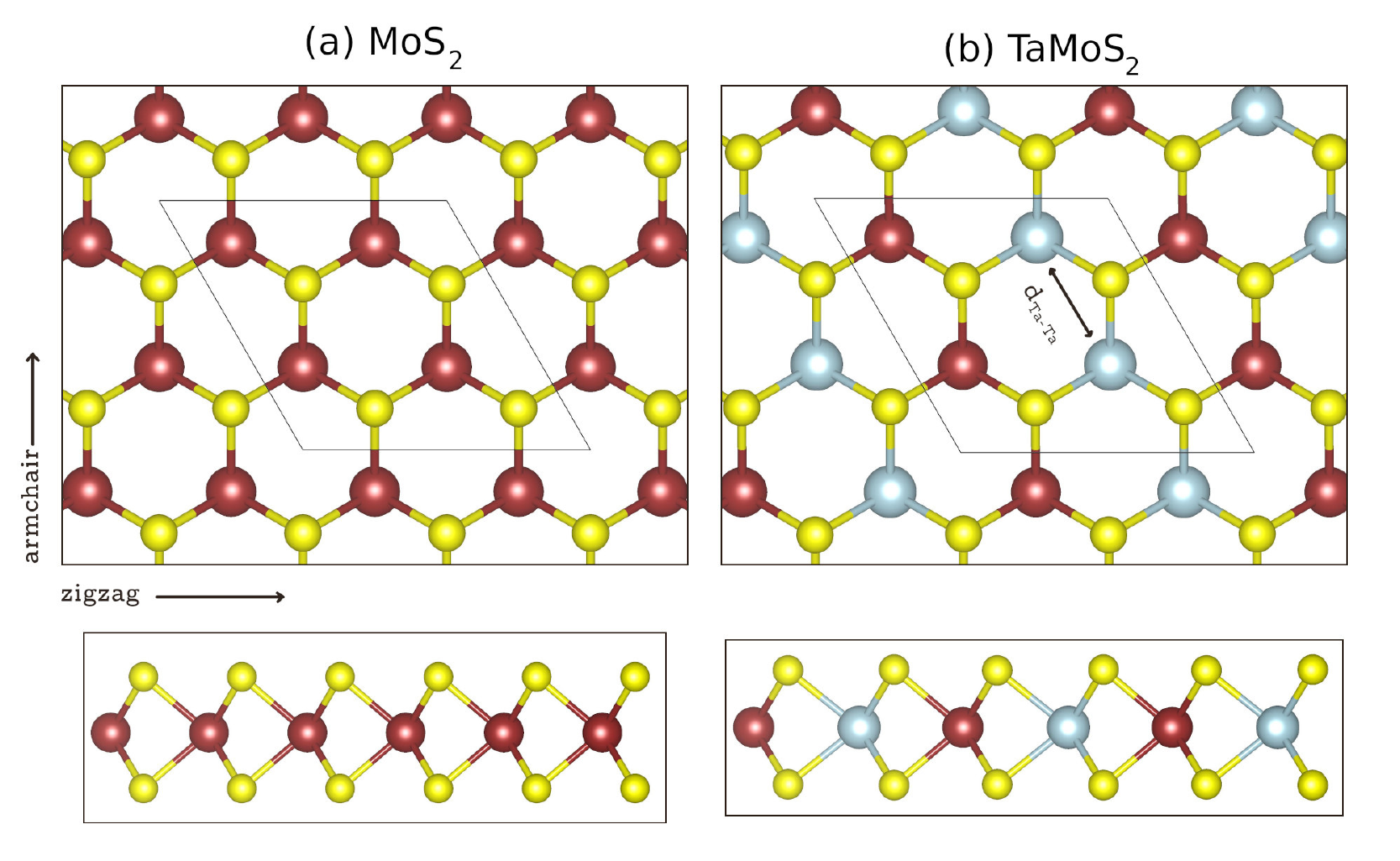}
	\caption{Pure MoS$_2$ monolayer (a), and Ta-doped MoS$_2$ monolayer (b) of both top view (top panels), and side view (bottom panels). The Mo, S and Ta atoms are colored brown, yellow, and cyan, respectively.}
	\label{fig01}
\end{figure}

The unit cell of a MoS$_2$ layer consists of 6 atoms; two molybdenum (Mo) atoms are located at the Wyckoff 2$c$ sites and four sulfur (S) atoms at the Wyckoff 4$f$ sites \cite{MOLINASANCHEZ2015554}. So a $2\times2$ supercell single layer of MoS$_2$ contains  
four Mo and eight S atoms as is shown in the parallelogram of \fig{fig01}(a).
The bonding type is mainly covalent within the atomically thin S-Mo-S layers.
In our calculation of MoS$_2$ monolayer, the obtained bond length between Mo-S atoms and the lattice constant are 
$2.41 \, \angstrom$, and $3.18 \, \angstrom$, respectively. These are very close to experimental values of Mo-S bond length, $2.383 \, \angstrom$, and lattice constant, $3.16 \, \angstrom$ \cite{MOLINASANCHEZ2015554}. The bond angles of the Mo-S-Mo and S-Mo-S in the pure MoS$_2$
monolayer are both $82.55^\circ$, with $81.55^\circ$ being theoretically obtained earlier \cite{doi:10.1021/acsomega.9b01429}. The separation between sulfur layers is $3.126 \, \angstrom$, which can be compared to the experimental value, $3.18 \, \angstrom$, measured by means of STM images.

In the Ta-doped MoS$_2$ system identified as the TaMoS$_2$ structure (see \fig{fig01}b), four atomic configurations based on the distance between the Ta atoms, $d_{\rm Ta\text{-}Ta}$, are considered. The concentration of the Ta dopant atoms is assumed to be $16.6 \%$ corresponding to two Ta atoms doped in the $2\times2$ supercell of the MoS$_2$ structure. The distance between the two Ta atoms are $2.56$, $3.01$, $4.14$, and $4.47 \, \angstrom$ in the four atomic configurations, identified as TaMoS$_2$-1, TaMoS$_2$-2, TaMoS$_2$-3, and TaMoS$_2$-4 structures, respectively. The two Ta atoms are substitutionally doped in the MoS$_2$ structures in such a way that one Ta is exchanged for a Mo atom while the other one replaces an S atom in all four atomic configurations. In this way, we can keep the semiconductor behavior of the TaMoS$_2$ with a finite bandgap. 

\begin{figure}[htb]
	\centering
	\includegraphics[width=0.5\textwidth]{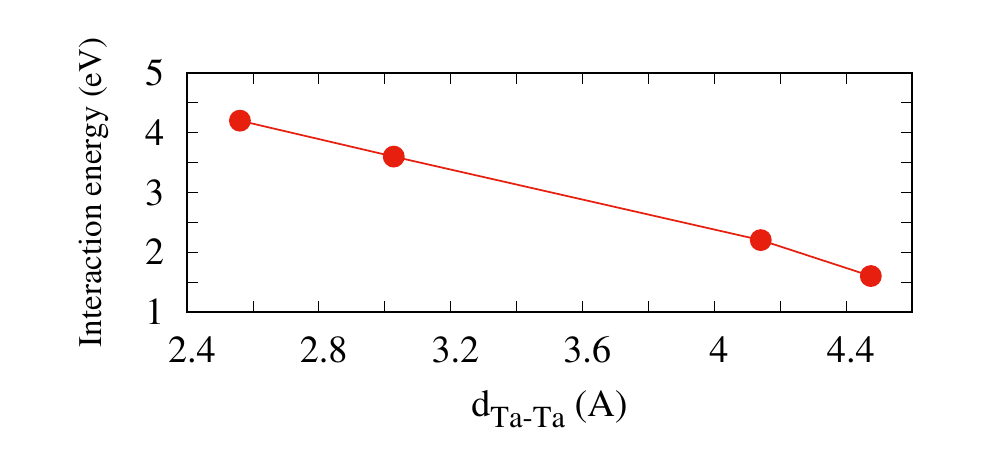}
	\caption{Interaction energy, $E_{\rm int}$, between the Ta atoms versus the distance between the Ta atoms, d$_{\rm Ta\text{-}Ta}$}
	\label{fig02}
\end{figure}

The four configurations of  TaMoS$_2$ with different distances between Ta atoms gives different interaction strength between the dopant atoms. The interaction energy can be defined as \cite{doi:10.1063/1.4742063, ABDULLAH2021106073}
\begin{equation}
	E_{\rm int} = E_2 - E_0 + 2 \times E_1,
\end{equation}
where $E_0$, $E_1$ and $E_2$ are the total energies of the systems with zero, one
and two substitutional dopant Ta atoms, respectively. We found that the interaction is repulsive as the interaction energy has a positive value for all the four considered configurations.
The repulsive energy is plotted as a function of the distance between the Ta atoms in \fig{fig02}. One can see that as the distance between the Ta atoms is increased the repulsive interaction between the dopant atoms is decreased.  TaMoS$_2$-1 has the strongest attractive interaction while TaMoS$_2$-4 has the weakest attractive interaction among the considered TaMoS$_2$ structures.

The attractive interaction between the Ta dopant atoms modifies the Mo-S, and Mo-Ta bond lengths, and the bond angles of Mo-S-Mo and S-Mo-S as is presented in \tab{table_one}. We can clearly see that the stronger repulsive interaction between Ta atoms increases the lattice constant of the TaMoS$_2$ structure. So, the TaMoS$_2$-1(TaMoS$_2$-4) has the longest(shortest) lattice constant among the TaMoS$_2$ structures. 
\begin{table}[h]
	\centering
	\begin{center}
		\caption{\label{table_one} Lattice constant, a, Mo-S-Mo bond angle, S-Mo-S bond angle, Mo-Ta, and Mo-S bond length for TaMoS$_2$ structures. The unit of all bond length is $\angstrom$.}
		\begin{tabular}{|l|l|l|l|l|l|l|}\hline
			Structure	  & a       & Mo-S-Mo  & S-Mo-S  & Mo-Ta   & Mo-S    \\ \hline
			TaMoS$_2$-1	  & 3.327   & 83.819   & 80.585  & 2.560   & 2.444   \\
			TaMoS$_2$-2	  & 3.261   & 83.583   & 82.976  & 2.560   & 2.420   \\
			TaMoS$_2$-3	  & 3.201   & 83.22    & 82.171  & 2.566   & 2.46    \\
			TaMoS$_2$-4	  & 3.192   & 77.736   & 83.144  & 2.99    & 2.5     \\   \hline
		\end{tabular}
	\end{center}
\end{table}

\begin{figure}[htb]
	\centering
	\includegraphics[width=0.45\textwidth]{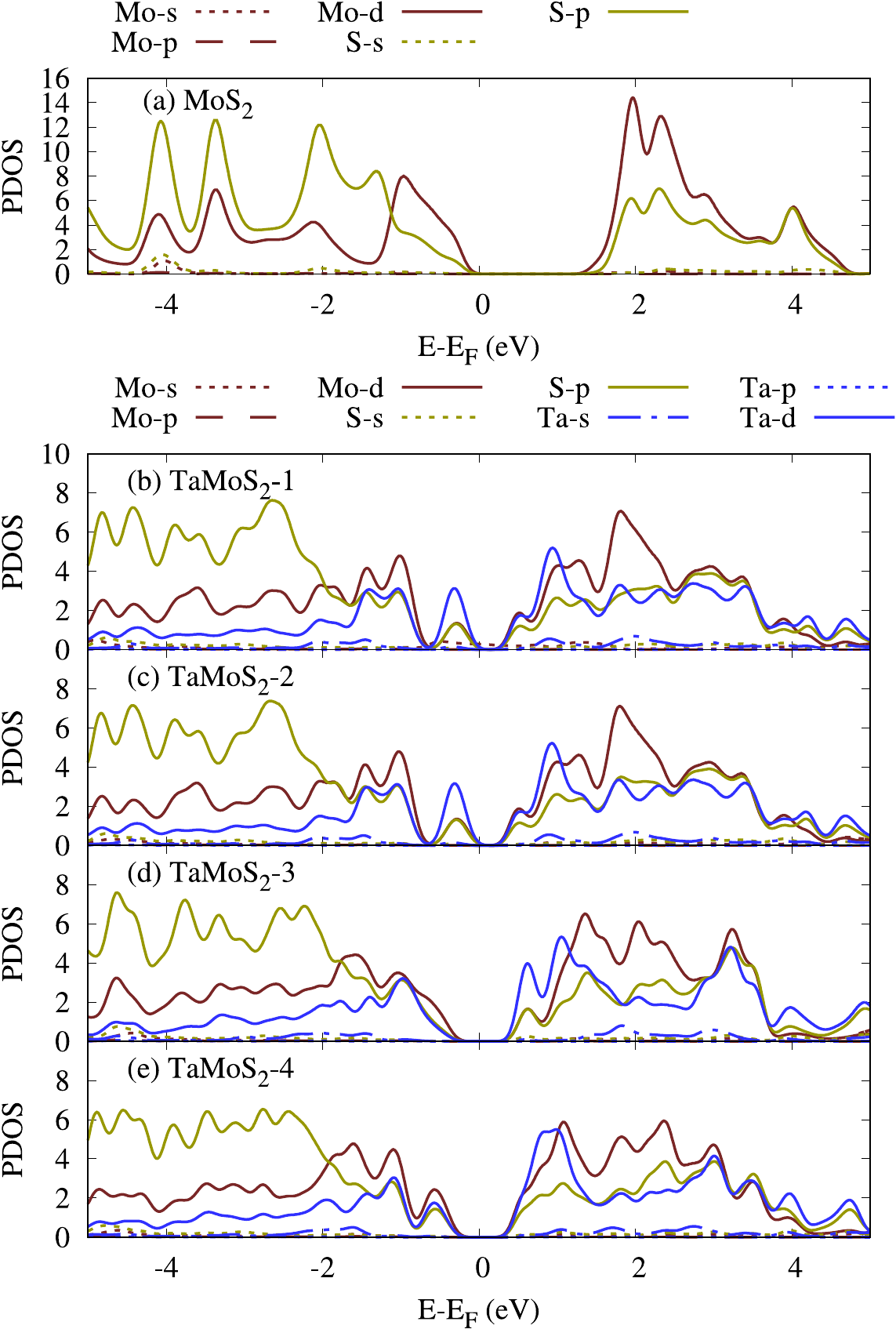}
	\caption{Partial density of states (PDOS) of the MoS$_2$ (a) and TaMoS$_2$ (b-e). The Fermi energy is set to zero.}
	\label{fig03}
\end{figure}

The Ta-Ta repulsive interaction elongates the unit cell of the TaMoS$_2$ causing a high sub-lattice symmetry breaking. It thus influences the density of states and the band structures of MoS$_2$. The partial density of state, PDOS, of the MoS$_2$ (a) and TaMoS$_2$ structures (b-e) are presented in \fig{fig03}. The major PDOS contribution in the pure MoS$_2$ monolayer around Fermi energy is the $d$-orbital of the Mo atoms (Mo-$d$) hybridized with the $p$-orbital of the S atoms (S-$p$).
The contributions of Mo become more relevant as the energy approaches the band gap because the atomic radius of an Mo atom is larger than that of an S atom.
In addition, states which come from the S-$p$ orbitals are, more importantly, 1 eV below the 
Fermi level.

The Mo-$s$ and the S-$s$ orbital contributions appear really deep in energy, they
are not expected to play an important role in this system due to their highly stable nature.
We therefore assume only Mo-$d$ and S-$p$ orbitals. Our PDOS analysis of MoS$_2$ is well in agreement with the literature \cite{MIRALRIO2018758}. 

In the presence of the Ta-Ta repulsive interaction, several new states are introduced near the Fermi energy. 
Also, a major Ta atoms rearrangement, relative to the pristine MoS$_2$ case, in the contributions of each element to the total DOS is apparent. 
In the presence of a strong Ta-Ta repulsive interaction, in TaMoS$_2$-1, a peak in the valence band region below the Fermi energy from $0$ to $-0.6$~eV and several peaks in the conduction band region from $0$ to $1.6$~eV due to the Ta-Ta repulsive interaction are seen. The contribution of Ta atoms in these peaks is decreases with decreasing Ta-Ta interaction strength. This is clear in TaMoS$_2$-4, especially in the valence band region. In contrast, with a decreasing Ta-Ta interaction strength the unoccupied states generated by the Ta atoms in the conduction band region is enhanced. 
Again, the $d$-orbitals of Ta atoms are the most active ones, Ta-$d$ (solid blue in Fig.\ \ref{fig03}). 
The contribution of the Ta atoms in the density of states decreases the PDOS of the Mo and S atoms. 

\begin{figure}[htb]
	\centering
	\includegraphics[width=0.4\textwidth]{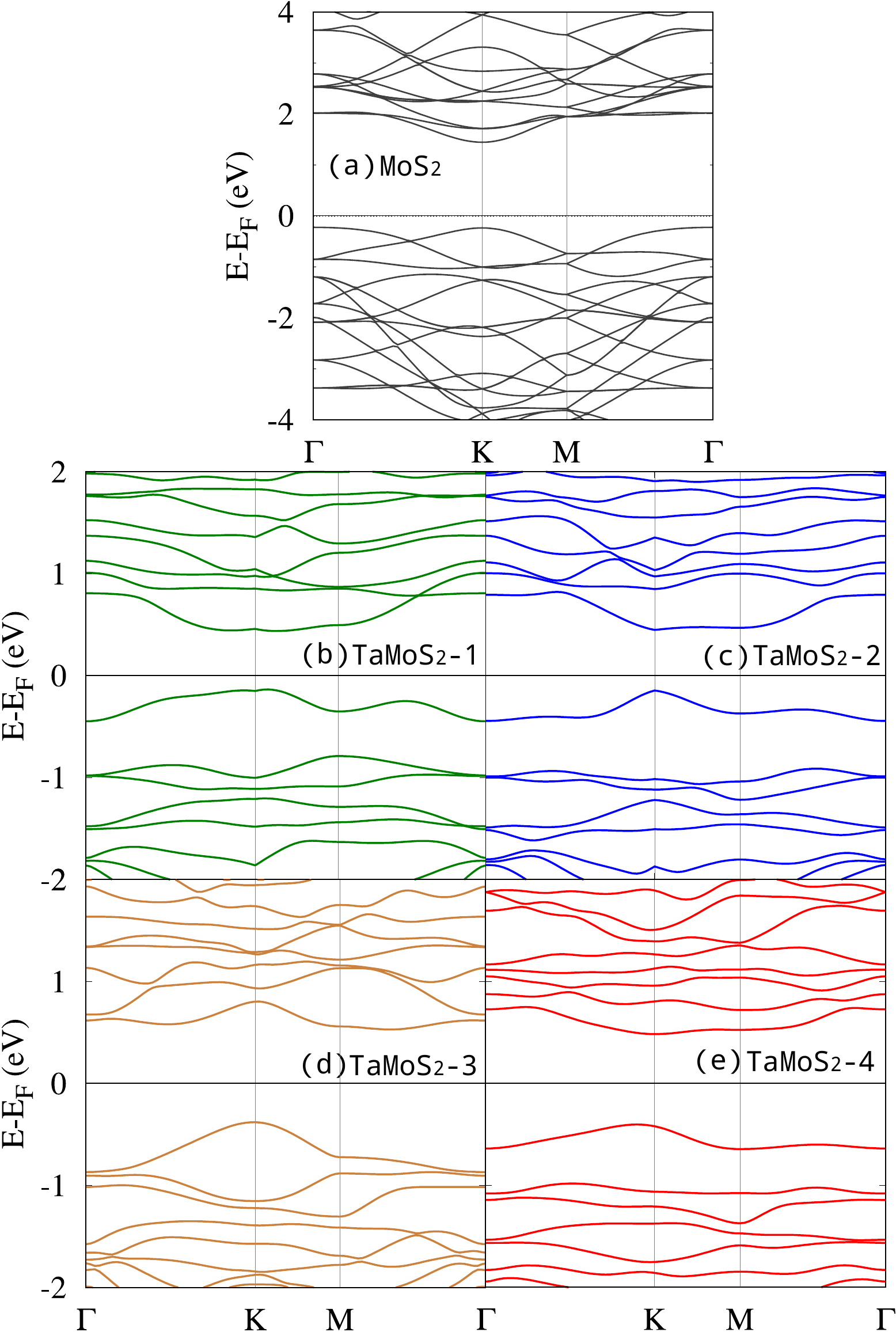}
	\caption{Band structure for optimized structures of the MoS$_2$ (a) and TaMoS$_2$ (b-e). The energies are with respect to the Fermi level, and the Fermi energy is set to zero.}
	\label{fig04}
\end{figure}

Another consequence of the levels introduced by the Ta-Ta interaction is the reduction of the band gap. The band structure of MoS$_2$ (a) and TaMoS$_2$ (b-e) are presented in \fig{fig04}. The band gap of pure MoS$_2$ monolayer is found to be direct and the band gap value is $1.68$~eV at the K point which is agree with the previous results \cite{Ryou2016}. As we have seen from the PDOS, the top of valence band and the bottom of conduction band are mainly formed by the $d$-orbital of the Mo atoms in MoS$_2$ monolayer.

\begin{figure}[htb]
	\centering
	\includegraphics[width=0.5\textwidth]{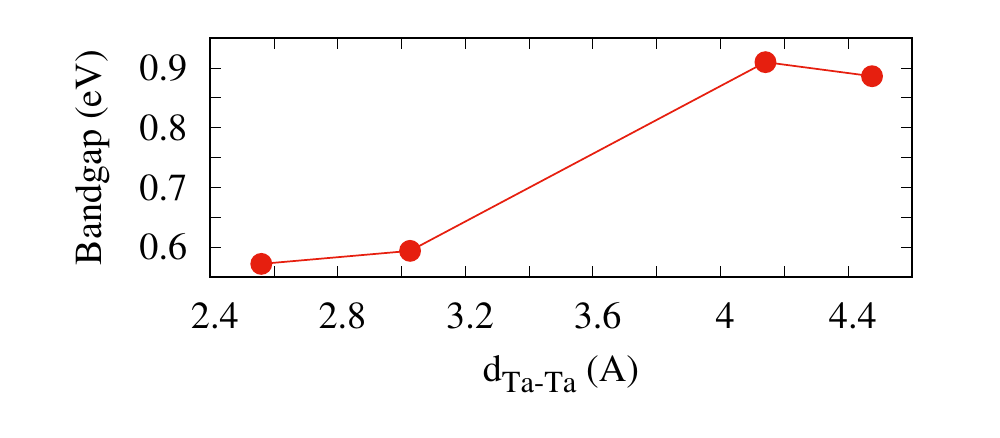}
	\caption{Band gap versus the distance between the Ta atoms, d$_{\rm Ta\text{-}Ta}$}
	\label{fig05}
\end{figure}

In the presence of the repulsive Ta-Ta interaction in TaMoS$_2$ the band gap is reduced as states are generated in the band gap region due to the $d$-orbitals of the Ta atoms. 
The stronger the repulsive interaction, the smaller band gap is seen, as can be seen from 
\fig{fig05}. The band gap values of TaMoS$_2$-1, TaMoS$_2$-2, TaMoS$_2$-3, and TaMoS$_2$-4 are 
$0.572$, $0.594$, $0.909$, $0.886$~eV, respectively. In addition, the degeneracy of the energy levels is reduced with decreasing Ta-Ta interaction strength in both the valence and the conduction band regions (see \fig{fig04}(d)).

The asymmetry of the band structure and the density of states caused by the Ta-Ta interaction is useful for controlling the thermoelectric properties of the systems \cite{ABDULLAH2020100740}. We are interested in the investigation of the thermoelectric properties in the temperature range
$T = 20\text{-}160$~K, where the electron contribution to the transport is dominant 
and the phonon participation can be neglected as the electron and the lattice temperatures
are decoupled in this temperature range \cite{PhysRevB.87.241411, ABDULLAH2020126578}.
\begin{figure}[htb]
	\centering
	\includegraphics[width=0.4\textwidth]{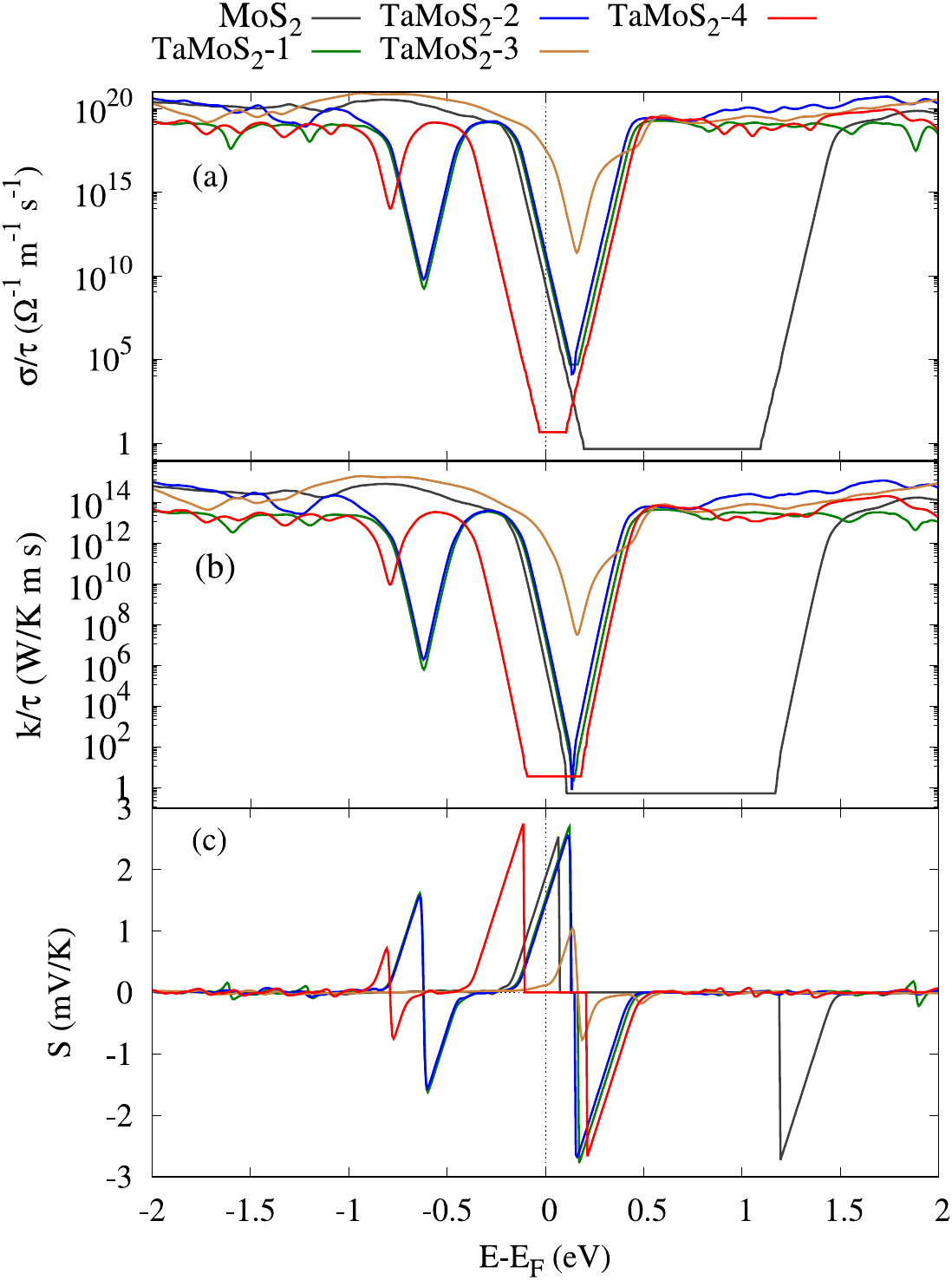}
	\caption{Electrical conductance, $\sigma$, (a), electronic part of thermal conductance, $k$, (b), and Seebeck coefficient, $S$, (c) versus energy for pure MoS$_2$ and TaMoS$_2$ monolayers. The Fermi energy is set to zero.}
	\label{fig06}
\end{figure}

We have used the BoltzTrap code to study the thermoelectric properties of the systems. 
The Boltzmann theory is implemented in the BoltzTrap code, which is utilized to obtain 
the semiclassical transport coefficients. The code
uses a mesh of band energies and is interfaced to the QE package.

\begin{figure}[htb]
	\centering
	\includegraphics[width=0.4\textwidth]{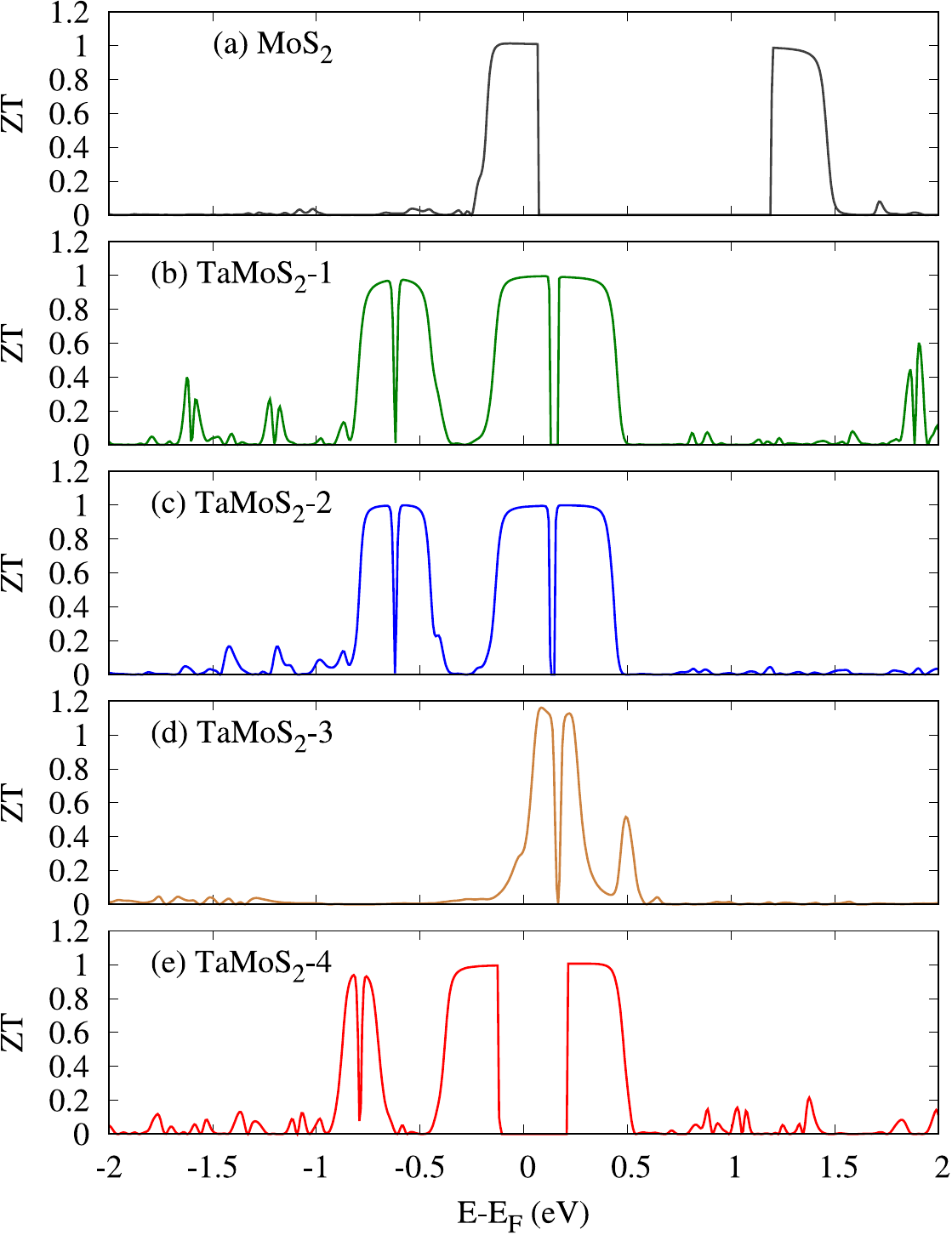}
	\caption{Figure of merit versus energy for pure MoS$_2$ (a) and TaMoS$_2$ (b-e) monolayers. The Fermi energy is set to zero.}
	\label{fig07}
\end{figure}

There are several attempt to enhance the figure of merit, $ZT$, which reflects the efficiency of thermoelectric devises. The $ZT$ can be defined as \cite{nano9020218, B822664B}
\begin{equation}
	ZT = \frac{S^2 \sigma}{k} T,
\end{equation}
where $S$ notes the Seebeck coefficient, $\sigma$ is the electrical conductance, $k$ is the thermal conductance, and $T$ is the temperature. In order to obtain a high $ZT$, one has to have high values of $S$ and $\sigma$, and a low value of $k$ \cite{ABDULLAH2021413273}. 
The $S$ (a), $\sigma$ (b), and $k$ (c) as functions of energy are plotted in \fig{fig06} for MoS$_2$ (a) and TaMoS$_2$ (b-e) monolayers. One can clearly see the influence of the wide band gap of MoS$_2$ on the thermoelectric properties. The values of $\sigma$, $k$, and $S$ are close to zero over a wide range of energy, due to the relatively large band gaps of MoS$_2$. It produces a vanishing $ZT$ over a wide range of energies from $0$ to $1.2$~eV as is shown in \fig{fig07}(a). We can thus confirm that the MoS$_2$ itself is not a good material for thermometric devices.  

In the presence of a repulsive Ta-Ta interaction, the enhanced symmetry of the density of states and the band structure enhance the Seebeck coefficient and thus a maximum value of $ZT$ is found for the 
TaMoS$_2$-3 in a small energy range around the Fermi energy. The value of $ZT \approx 1.2$ for TaMoS$_2$-3 is seen (see \fig{fig07}(d)). This results may be useful for thermoelectric devices.

\section{Conclusion and Remarks}\label{Sec:Conclusion}

The framework of this study is based on the Kohn-Sham density functional approach to investigate electronic and thermal properties of the Tantalum doped Molybdenum disulfide monolayer.
We found that the MoS$_2$ monolayer is not a good material for thermoelectric devices because of its wide band gap. To enhance 
its thermal behavior Ta atoms are used as dopant atoms. More precisely, we focus on the interaction effect between the Ta atoms doped in MoS$_2$ in this study. In the presence of a strong attractive interaction between Ta atoms, several new states are introduced near the Fermi energy. Consequently, the band gap is reduced in the still semiconducting TaMoS$_2$. The reduced band gap can enhance the Seebeck coefficient and the figure of merit. 
In the presence of a weak Ta-Ta attractive interaction, the band gap is slightly reduced but remains large, 
which is promising for thermoelectric devices. We thus see vanishing Seebeck coefficient and figure of merit over a wide range of energy. Our study may be useful for thermoelectric devices such as thermoelectric generators.

\section{Acknowledgment}
This work was financially supported by the University of Sulaimani and 
the Research center of Komar University of Science and Technology. 
The computations were performed on resources provided by the Division of Computational 
Nanoscience at the University of Sulaimani.  



\end{document}